\documentclass[12pt]{iopart}

\usepackage{cite}
\usepackage{graphicx}

\eqnobysec

\begin{document}

\title[Interacting semi-flexible self-avoiding walks]{Interacting semi-flexible self-avoiding walks  studied on a fractal  lattice}

\author{Du\v{s}anka  Mar\v{c}eti\'{c}}

\address {University of Banja Luka, Faculty of Natural Sciences and Mathematics,  M.~Stojanovi\'{c}a 2,  78 000 Banja Luka, Bosnia and Herzegovina}
\ead{dusanka.marcetic@pmf.unibl.org}

\begin{abstract}
 Self-avoiding   walks are studied on the  3-simplex fractal lattice as  a  model of  linear polymer conformations  in a dilute, non-homogeneous solution.  A model is supplemented with bending energies and attractive-interaction energies   between  non-consecutively  visited  pairs of nearest-neighboring  sites (contacts).  It captures the  main features of a  semi-flexible polymer subjected  to variable  solvent conditions. Hierarchical structure of the fractal lattice enabled  determination of the exact recurrence equations for the generating function, through which  universal and local properties of the model  were studied. Analysis of the recurrence equations  showed that  for all finite values of the considered  energies  and non-zero temperatures,  polymer  resides in an expanded  phase.     Critical exponents  of the  expanded phase are universal and  the    same as those for ordinary self-avoiding  walks on the same lattice found earlier. As a measure of local properties,  the  mean number of contacts per mean number of steps as well as persistence length, are calculated as functions of   Boltzmann weights associated with  bending energies and attractive interactions between contacts.  Both quantities are monotonic functions of stiffness weights for fixed interaction,  and   in the limit of infinite stiffness  the  number of contacts  decreases to zero,  while  the persistence length increases   unboundedly.
\end{abstract}

%
%
%
%
%
\vskip 5mm


\maketitle
\section{Introduction}
\label{intro}

 A self-avoiding walk  (SAW) on a lattice is a sequence of consecutive  steps along the lattice bonds  in which the steps are not allowed to go through  the same lattice site  more than once~\cite{ Mad}.   Due to their vast range of applications in different areas, but  most commonly  in polymer related problems~\cite{Gut,Ren}, SAWs   have become subject of continuous  research interest since their introduction~\cite{Or,Flori}.  They  belong to the class  of   combinatorial concepts  which are simple to state but difficult or even impossible to solve exactly.    In polymer physics, SAWs are canonical lattice model for      conformations of a single  linear  polymer  dissolved   in a good solvent.  Non-intersecting property of  SAWs  corresponds  to an excluded volume effect of a  polymer, and that is the only  interaction relevant  in  a good solvent conditions, i.e. at high temperatures. More realistic model that captures the  behavior of a polymer  at all solvent temperatures can be obtained by implementing  ordinary SAWs  with an attractive interaction between contacts, i.e. visited pairs of sites that are nearest-neighbors but not adjacent along the walk. Such an interaction  represents monomer-monomer interaction mediated by the solvent and thus mimics different solvent qualities. Many aspects of the interacting model have been intensively  studied  concerning the polymer  collapse transition~\cite{Fis, Ish1,  Ish2, Ish3, Der, Sal,  Bin, Nem, Gras, Doug, Bar, Nid,Tesi2,Ben, Fost, Vog,Pon,Beat}. Another step that improves the model  is the addition of bending energy to each bend in the walk so that  various degrees of stiffness of natural polymers could  be accounted for. Influence  of stiffness on the polymer collapse and phase diagram in general  has also been considered  on regular  lattices~\cite{Kol,Bas, Doye, Lise,Kraw}. In all of these studies it is assumed that  the polymer is immersed in a homogeneous solution.  However, in many real situations that might  not  be the case if some impurities   or obstacles are   dispersed in solution.  Presence of disorder  can affect the  critical properties of polymers modelled by the SAWs~\cite{Chak,Bred}.   In order to study its impact, disordered environment     is  usually represented by    translationally non-invariant  lattices such us randomly diluted and fractal lattices.  Deterministic  fractals have an advantage of being   scale invariant,  a symmetry   that allows for many exactly solved problems by the application of an exact recursive method. Critical behavior of  self-avoiding walks with  contact interactions or bending stiffness  on fractal lattices are studied  in~\cite{Klein, Dhar1, Knez1, Kumar, Knez2,Ziv0, Gia}. Despite  aforementioned advantage of fractal lattices,  SAWs with  both, interaction and stiffness,  require  in  their recursive treatment  a large  number of variables  even on  the simplest fractal lattice.  Therefore, to the best of our knowledge, there is only one such study~\cite{Tut},  where the  phase diagram  on the 4-simplex lattice is obtained  by the   renormalization group approach.    \par

In the present paper, we  have considered self-avoiding  walks with  contact interactions and bending stiffness  (Interacting semi-flexible self-avoiding walk model, ISFSAW)  on the fractal, 3-simplex lattice.  Critical behaviour  of the  reduced models: Interacting self-avoiding walks (ISAW)~\cite{Dhar1},  Semi-flexible self-avoiding walks (SFSAW)~\cite{Gia} and  ordinary Self-avoiding walks (SAW)~\cite{Dar2}  have already been studied on this lattice with the focus on critical  exponents. Here   we found  that  global behavior of  the compound  model is almost the same as  the behavior of the reduced models, since    all  finite values of the interaction parameters  describe a  polymer  in an expanded phase.  At zero temperature,  flexible polymer is in a  compact phase, while semi-flexible polymer  is in a rigid-rod phase.   Furthermore,  we have studied nonuniversal properties of the model.   In the framework of the grand canonical ensemble, we have determined critical  fugacity  surface and studied  the influence of    bending and interaction energies on the  structure of conformations  by calculating  the mean number of contacts per mean number of steps and persistence length. In  section~\ref{druga},  we describe the model, outline a  method in general and   apply it on the 3-simplex lattice. Analysis of the recurrence equations  and calculation procedure are presented in section~\ref{treca}, while results and discussion  are presented in subsection~\ref{rez1} for flexible  and subsection~\ref{rez2} for semi-flexible walks.   Summary of the paper and conclusions are  presented  in  section~\ref{cetvrta}.  Appendix A contains a  complete  set of recurrence equations complementing  the generating function  of open walks.

\section{ A model and  grand canonical formalism  on the 3-simplex lattice}\label{druga}

\begin{figure}[t]
\begin{center}
\includegraphics[scale=0.9]{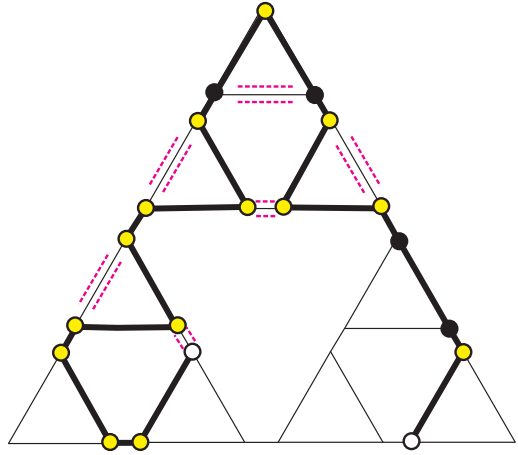}
\end{center}
\caption{A self-avoiding walk   consisting of  $N=19$ steps   on  the third order generator of the 3-simplex lattice.  Lattice sites visited by the walk are represented with circles in different colors, where  white  circles represent sites with the  endpoints of the walk.  At each yellow site the walk  makes a bend to which a   Boltzmann   factor $s$ is assigned. Attractive interactions between contacts are marked with the double dashed lines and weighted  with the  factors $u$. Finally, each step of the walk is weighted with a fugacity $x$, so that the  total weight of the walk  is $x^{19}s^{14}u^{6}$. }
 \label{fig:saw}
\end{figure}

 The 3-simplex  lattice is a  finitely ramified  fractal lattice  introduced by Dhar~\cite{Dar2, Dhar3}.  The lattice  is similar to  the Sierpinski gasket and   has the same fractal dimension $d_f=\frac{\ln 3}{\ln 2}$, but   is  less demanding  regarding  the number of variables and their equations in  studying  SAW lattice problems~\cite{Ram}.  Lattice construction follows the iterative steps, where   prefractal  structure obtained in the   $r$-th step  is called the  $r$-th order generator ($G^{(r)}$), and the whole lattice is obtained by letting $r\to\infty$.  The  third order generator  is shown  in~\Fref{fig:saw} altogether with one  self-avoiding walk consisting of  $N=19$ steps. In the ISFSAW model each polymer conformation is represented by the  weighted walk on a lattice:    Boltzmann factors  $u=\exp(-\epsilon/kT)$    and    $s=\exp(-\epsilon_b/kT)$  are associated with  each contact and each bend of the walk, respectively.  These factors  are  illustrated  along with the  walk shown   in \Fref{fig:saw}.   Interactions that we consider here are restricted  to   $-\infty\leq\epsilon\leq 0$  and $0\leq\epsilon_b\leq\infty$, but these sets   can be extended to include super-perfect solvent ($\epsilon> 0$)~\cite{Fis}  and super-flexible walks $\epsilon_b<0$~\cite{ Pol,Ziv}.
 Energy of  an $N$-step walk with  $M$ contacts and $N_b$ bends is $E_N(M,N_b)=M\epsilon+N_b\epsilon_b$,  so that canonical partition function is  $Z_N=\sum_{N_b}\sum_{M} W_N(N_b,M)s^{N_b}u^{M}$, where   $W_N(N_b,M)$ is the number of  different $N$-step self-avoiding  walks    of  energy   $E_N(M,N_b)$, per lattice site.
Assigning   a  fugacity $x$ to   each step of the walk,  grand canonical partition function reads  $\mathcal{G}(x,s,u)=\sum_{N=0}^{\infty}\sum _{N_b}\sum_{M}W_N(N_b,M)x^Ns^{N_b}u^{M}$.  It is an overall  weight of all possible walks, and the weight  $x^{19}s^{14}u^{6}$ of the  walk shown in \Fref{fig:saw} is just one term in $\mathcal{G}$.  Grand canonical partition function  is   the generating function for  the sequence $Z_N$  and can also be written as    $\mathcal{G}=\sum_{N=0}^{\infty}x^NZ_N$. It is assumed that  the  leading order behaviour of  $Z_N$   is $Z_N\sim\omega(s,u)^N$ in the limit $N\to\infty$,  so that  the radius of convergence of    $\mathcal{G}$ is given by $x_c(s,u)=(\omega(s,u))^{-1}$.  For $x<x_c$ the mean length  $\langle N\rangle$ of walks is  finite,  whereas   for   $x\geq x_c$  it diverges.  The so called  polymerization transition, which occurs at  critical fugacity  $x_c$,       corresponds  to  the thermodynamic limit.  Free energy per step   in the thermodynamic limit,  defined as  $f=-k_BT\lim_{N \to \infty}(\ln Z_N/N)$, is then  given  by  $f=-k_BT\ln\omega(u,s)=k_BT\ln x_c(u,s)$, from which all  thermal, bulk properties of the system can be found. Here, it is  more convenient  to work in the grand canonical formalism, where  the  mean number of steps,  contacts and bends  are  given by
\begin{equation}\label{eq:meanN}
  \langle N\rangle =\frac{1}{\mathcal{G}}\sum_{N, M,N_b}N W_N(N_b,M)x^Ns^{N_b}u^{M}= \frac{x}{\mathcal{G}}\frac{\partial \mathcal{G}}{\partial x}\,,
\end{equation}
\begin{equation}\label{eq:meanM}
  \langle M\rangle =\frac{1}{\mathcal{G}}\sum_{N,M,N_b}M W_N(N_b,M)x^Ns^{N_b}u^{M}=\frac{u}{\mathcal{G}}\frac{\partial \mathcal{G}}{\partial u}\,,
\end{equation}

\begin{equation}\label{eq:meanNb}
  \langle N_b\rangle =\frac{1}{\mathcal{G}}\sum_{N,M,N_b}N_b W_N(N_b,M)x^Ns^{N_b}u^{M}=\frac{s}{\mathcal{G}}\frac{\partial \mathcal{G}}{\partial s}\,.
\end{equation}
 Quantities of interest: the mean number of contacts and  the mean number of bends per mean number of steps, are

 \begin{equation}\label{eq:meanm}
     m=\frac{\langle M\rangle}{\langle N\rangle}=\frac{u}{x}\frac{\mathcal{G}^{u}}{\mathcal{G}^{x}}\qquad
     \mbox{and}  \qquad  n_b=\frac{\langle N_b\rangle}{\langle N\rangle}=\frac{s}{x}\frac{\mathcal{G}^{s}}{\mathcal{G}^{x}}\,,
 \end{equation}
where $\mathcal{G}^{x}=\left(\frac{\partial \mathcal{G}}{\partial x}\right)_{u,s}$,  $\mathcal{G}^{u}=\left(\frac{\partial \mathcal{G}}{\partial u}\right)_{x,s}$  and $\mathcal{G}^{s}=\left(\frac{\partial \mathcal{G}}{\partial s}\right)_{x,u}$. All  partial derivatives should be   calculated at the critical value $x_c(u,s)$. A measure of polymer persistency obtained as    $l_p=\langle N\rangle/\langle N_b\rangle=1/n_b$, has the meaning of an  average  number of steps between two consecutive bends (average length of straight segments) and  will  be considered  as a persistence length throughout this paper.

    \par
\begin{figure}[t]
\begin{center}
\includegraphics[scale=0.68]{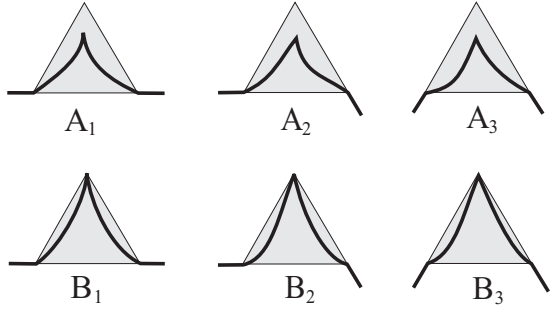}
\end{center}
\vspace{-0.2cm}
\caption{Restricted generating functions  necessary for   recursive formulation of the  ISFSAW model on the 3-simplex lattice.  They are classified according to the number of visited corner vertices ($A$ and $B$) and  directions of the first  steps  external to the generator ($1$, $2$ and $3$). Also, $A_i\supset B_i$ for $i=1,2,3$.}
\label{fig:pomocne}
\end{figure}
\begin{figure}[t]
\begin{center}
\includegraphics[scale=0.85]{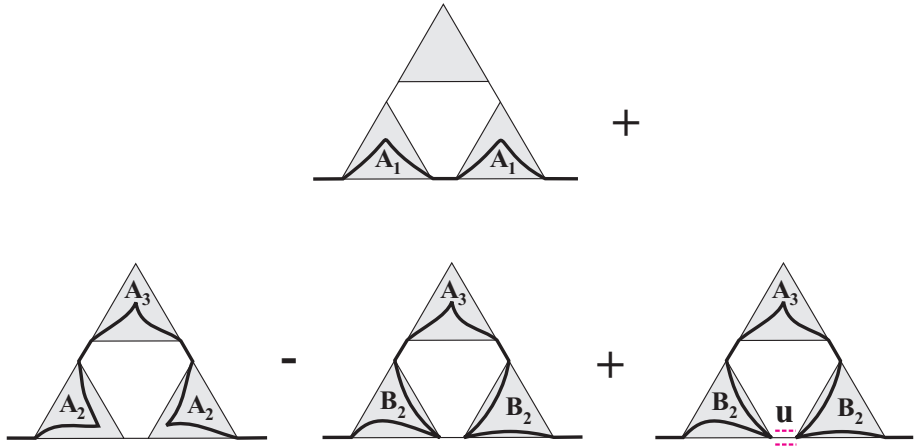}
\end{center}
\caption{ Schematic representation of the procedure   with which    recurrence equation (\ref{eq:re1a}) is obtained. Larger triangle is a generator of order $r+1$, while  smaller, shaded triangles are  the generators of the $r$-th order.  The overall weight of the walk  on the  larger triangle is obtained  by the multiplication of the  weights  assigned to   the smaller constitutive triangles. Each such overall weight corresponds to one term in recurrence equation (\ref{eq:re1a}).  The third term is subtracted from the second   and added multiplied by the factor $u$ because   an interaction between smaller triangles occured. }
\label{fig:rekA}
\end{figure}

In order to construct $\mathcal{G}(x,s,u)$ for the ISFSAW model  on  the 3-simplex lattice, one finds that  twenty  restricted generating functions are   involved, but  it turned out that only six of them are actually strictly necessary. These six    restricted generating  functions are  the weights of the corresponding  SAWs, denoted as $A_1$, $A_2$, $A_3$, $B_1$, $B_2$ and $B_3$,   which  comprise self-avoiding  polygons, or walks (open) with the  endpoints fixed at the corner vertices of the largest  generator.   They are   shown schematically in \Fref{fig:pomocne}.  With $A$  we denote all  SAWs   that enter a  generator (of any order)  at one corner vertex and leave it at another.  $A$-type walks  includes both:   walks that visit the  third corner vertex of a generator  and those which do not visit it. In order to  properly account for the   interactions at all levels of the fractal structure, $B$-type of  walks are introduced as   those   subset of walks $A$   which obligatory visit the  third  corner vertex of the generator. With that set of  variables,  recurrence equations of the interacting model  can be obtained  from  the equations for  ordinary SAWs by subtracting  some terms with weights  $B$  from  weights $A$, and  adding them    properly with the interaction~\cite{Dar2, Dhar1}. Further classification to  $A_i$ and  $B_i$,  $i=1,2,3$,  is made  according to the direction of  external steps with which walks enter and leave  a   generator, in order to include possible bends at its   corner vertices.  Following the illustration   in \Fref{fig:rekA}  from which the   recurrence equation for variable $A_1$  is derived,  all  recurrence equations can be written as
\numparts
 \begin{eqnarray}
   A_{1\,r+1} &=A_{1\,r}^2+A_{2\,r}^2A_{3\,r}+(u-1) A_{3\,r}B_{2\,r}^2\,, \label{eq:re1a} \\
A_{2\,r+1} &=A_{1\,r}A_{2\,r}+A_{1\,r}A_{2\,r}A_{3\,r}+(u-1) A_{3\,r}B_{1\,r}B_{2\,r}\,, \label{eq:re1b} \\
   A_{3\,r+1} &=A_{2\,r}^2+A_{1\,r}^2A_{3\,r}+(u-1) A_{3\,r}B_{1\,r}^2\,, \label{eq:re1c}\\
   B_{1\,r+1} &=A_{2\,r}^2B_{3\,r}+(u-1) B_{2\,r}^2B_{3\,r} \,, \label{eq:re1d}\\
   B_{2\,r+1} &= A_{1\,r}A_{2\,r}B_{3\,r}+(u-1) B_{1\,r}B_{2\,r}B_{3\,r}\,,  \label{eq:re1e}\\
   B_{3\,r+1} &= A_{1\,r}^2B_{3\,r}+(u-1) B_{1\,r}^2B_{3\,r} \label{eq:re1f} \,.
\end{eqnarray}
\endnumparts
 Initial values given on a unit triangle are: $A_{1\,1}=x^2+x^3s^3u $, $A_{2\,1}=x^2s+x^3s^2u $, $A_{3\,1}=x^2s^2+x^3su$, $B_{1\,1}=x^3s^3u$, $B_{2\,1}=x^3s^2u$ and $B_{3\,1}=x^3su$.   Interaction and bending  parameters vary over the intervals  $[1,\infty]$ and  $[0,1]$, respectively. For $u=1$ and   $s=1$,  SAWs are non-interacting and flexible (ordinary SAWs).   Smaller $s$ corresponds to stiffer chain  and  $s\to 0$  is   a  rigid-rod limit.\par

\section{ Analysis of the model and calculation }\label{treca}
 Polygon generating function  is given  by
\begin{equation}\label{eq:genP}
    P(x,s,u)=\frac{1}{3}x^3s^3+\sum_{r=1}^{\infty}\frac{1}{3^{r+1}}\bigg(A_{3r}(x,s,u)\bigg)^3\,,
\end{equation}
  and can   also   be written  recurrently  in the form $P_{r+1}=P_r+\left(A_{3r}\right)^3/3^{r+1}$ with $P_1=(1/3)x^3s^3$. Iterating~(\ref{eq:genP})  together  with recurrence  equations~(\ref{eq:re1a})-(\ref{eq:re1f}),  for each particular pair of $u$ and $s$, critical fugacity $x_c$ is determined  as the largest value of $x$  at  which the  generating function $P$  converges. Such a behaviour of $P$   can be deduced  from  the behaviour of  the reduced generating functions $A_i$ and $B_i$, $i=1,2,3$.  Indeed, for every  $x<x_c(s,u)$  all the variables $A_i$ and $B_i$  tend toward zero, while for $x>x_c(s,u)$ they diverge (an so does $P$). For $x=x_c(s,u)$ they tend  to   $A^*, A^*,A^*,0,0,0$, with $A^*=0.61803...$.   This is the fixed point of the  system of nonlinear  difference equations~(\ref{eq:re1a})-(\ref{eq:re1f}), as can be checked.  By putting $B_i^*=0$ and $A_i=A^*$ for $i=1,2,3$,     the  fixed point  equation for variables $A_i$ reduces to $A^*={A^*}^2+{A^*}^3$ with the nontrivial  solution  $A^*=(\sqrt 5-1)/2=0.61803...$,  the same as in~\cite{Dhar1, Gia, Dar2}. This  nontrivial fixed point  corresponds to an expanded, swollen phase of a polymer,  with the  end-to-end critical exponent $\nu$ given by
   $\nu=\ln 2/\ln(\frac{7-\sqrt5}{2})\approx0.7986$~\cite{Dhar1, Dar2}.  Thus, for all finite and non-zero values of the parameters  $u$  and $s$,  polymer is in  the expanded phase with  constant  critical exponent $\nu$.  There is no collapsed phase on  the $3$-simplex  lattice for  any finite $u$.   Another  nontrivial  fixed point   $(1,0,0,0,0,0)$ of the system~(\ref{eq:re1a})-(\ref{eq:re1f})  can be  reached  for $s=0$ and $x_c=1$. It is an  unstable fixed point  that corresponds to a  rigid-rod phase of a polymer, also found in~\cite{Gia}. In this phase polymer is fully elongated, with no bends, and the   end-to-end distance scales with the number of steps  as $\mathcal{R}\sim N$, so that  $\nu=1$.
  Critical exponent $\nu$  can be obtained numerically by noticing that $\mathcal{R}_r\sim 2^r\sim N_r^{\nu}$,  where   $\mathcal{R}_r$ is a  root mean square  end-to-end distance of a walk  that occupies $r$-th order generator of linear size $2^r$,  and  $N_r$ is the number of steps of  the walk.  Writing similar   expression that holds  at the stage $r+1$, it  follows that $\nu=\ln(2)/(\ln N_{r+1}/\ln N_r)$. For all finite values of $u$ and  $s\neq0$, we have  numerically  obtained  $\nu\approx0.7986$.  When   $s=0$  then  $x_c$=1 and  $\nu=1$ numerically, but  it also  follows from the previous expression for $\nu$  given that $N_r\sim2^r$ in the rigid-rod phase.
    \par
Generating functions for polygons and open walks ($\mathcal{G}$) have the same  radius of convergence $x_c$,  but they behave differently at $x_c$.  Namely,  $P$ converges at $x_c$ while $\mathcal{G}$ diverges. Moreover, the mean length    of polygons (determined by the first derivative of $P$)   stays  finite  at $x=x_c$ and is short, so that thermodynamic limit is not   approached  even after many iterations. Therefore, we have calculated the  mean length $\langle N\rangle$  of open walks by using   the generating function
  \begin{equation}\label{eq:genG}
    \mathcal{G}(x,s,u)=x(1+xsu)+\sum_{r=1}^{\infty}\frac{1}{3^r}\Big(C_r^2+A_{3r}C_r^2+(u-1)A_{3r}D_{1r}^2+A_{3r}^2G_{3r}\Big),
\end{equation}
 with which  $\langle N\rangle$ diverges  at $x_c$ when the number of iterations tends toward $\infty$.   In  $\mathcal{G}$,  walks  with one or both endpoints in the interior vertices are  also contained.  Their recurrence equations are given in Appendix A.  But, as we have already mentioned, equations    (\ref{eq:re1a})-(\ref{eq:re1f}) suffice for the calculation of the  mean values (\ref{eq:meanm}), since they can be calculated by using the corner-to-corner restricted generating function $A_1$ which   behaviour at $x_c$  is the same as of $\mathcal{G}$. We have checked  that   all quantities  defined per step of the walk or per lattice site,  calculated via  $A_1$,  are the same as those calculated via  $\mathcal{G}$.
  To calculate  (\ref{eq:meanN})-(\ref{eq:meanm}), we need all  variables  $A_i$ and $B_i$, $i=1,2,3$,  and their  first derivatives with respect to $x$, $u$ and $s$.  From recurrence equations  (\ref{eq:re1a})-(\ref{eq:re1f})  we see that $A_{i\,r+1}=f_{i}\big(A_{j\,r}(x,u,s), B_{j\,r}(x,u,s),u\big)$  and $B_{i\,r+1}=g_{i}\big(A_{j\,r}(x,u,s), B_{j\,r}(x,u,s),u\big)$, where   $f_i$ and $g_i$ are some   polynomial functions  and   $i,j=1,2,3$.   With  variables that denote partial derivatives with respect to  $x$ holding  $u$ and $s$   constant:  $A^x_{i\,r+1}=\frac{\partial A_{i\,r+1}}{\partial x}$,
$B^x_{i\,r+1}=\frac{\partial B_{i\,r+1}}{\partial x}$, $A^x_{i\,r}=\frac{\partial A_{i\,r}}{\partial x}$ and $B^x_{i\,r}=\frac{\partial B_{i\,r}}{\partial x}$,   recurrence equations
for partial derivatives   can be written in the form of a  matrix equation,  where $ 6\times6$ matrix connects two   columns of order $6\times1$:
\begin{equation}\label{eq:reizvodi}
\left(
  \begin{array}{c}
    A^x_{i\,r+1} \\
    B^x_{i\,r+1} \\
  \end{array}
\right)=\left(
          \begin{array}{ccc}
           \frac{\partial f_{i}}{\partial A_{j\,r}}  &   \frac{\partial f_{i}}{\partial B_{j\,r}}  \\
            \frac{\partial g_{i}}{\partial A_{j\,r}}  &   \frac{\partial g_{i}}{\partial B_{j\,r}}  \\
          \end{array}
        \right)\left(
                 \begin{array}{c}
                 A^x_{j\,r} \\
                  B^x_{j\,r} \\
                 \end{array}
               \right)\,\quad  $i,j=1,2,3$.
\end{equation}
 Initial conditions  are obtained from the corresponding initial conditions for variables $A_i$ and $B_i$, and  are given by  $A^x_{1\,1}=2x+3x^2s^3u$, $A^x_{2\,1}=2xs+3x^2s^2u$, $A^x_{3\,1}=2xs^2+3x^2su$, $B^x_{1\,1}=3x^2s^3u$, $B^x_{2\,1}=3x^2s^2u$ and $B^x_{3\,1}=3x^2su$. In a similar manner we obtained  recurrence equations for partial derivatives with respect to $s$ and $u$. Thus,  for each    $s$, $u$  and $x_c(s,u)$,  six original  and eighteen additional recurrence equations (for first derivatives) were iterated,  and quantities   (\ref{eq:meanN})-(\ref{eq:meanm}) calculated. One should mention  that all results obtained  for $m$ and $l_p$ that we present    in the  next subsections  should hold equally  well for walks and polygons, since  both types of walks have the same growth constant $\omega(s,u)=1/x_c(s,u)$  and thus the  free energy in the thermodynamic limit.   Equality of $m$ for  walks and polygons has  been  deduced  and numerically confirmed on  regular lattices as well~\cite{Tesi2,Ben}.

\subsection{  Results and discussion, $s=1$} \label{rez1}
In this subsection we present results for critical fugacity, density of monomers, $m$ and $l_p$  for flexible interacting  walks ($s=1$, $1\leq u\leq\infty$).

\begin{figure}[t]
\begin{center}
\includegraphics[scale=0.45]{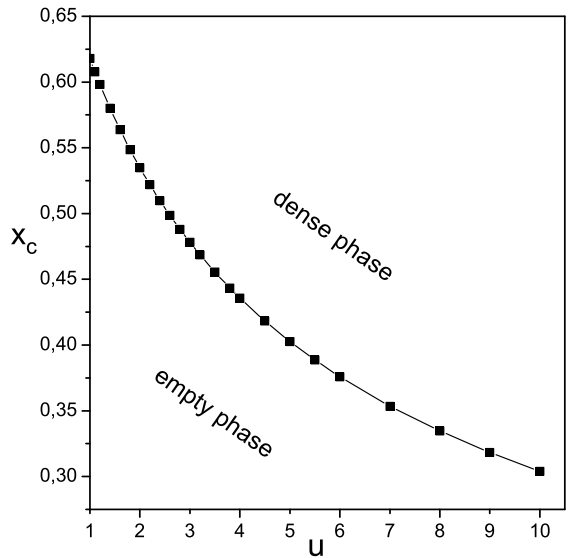}
\end{center}
\vspace{-0.4cm}
\caption{Dependence of critical fugacity $x_c$ on the  interaction parameter $u$ for $s=1$. Above the critical line, the phase is polymerized with the finite density of monomers (dense),  whereas below the critical line it  is unpolymerized with zero density (empty). At the critical line, the phase is the  polymerized, expanded  with zero density. }
\label{fig:fugx}
\end{figure}
 In \Fref{fig:fugx}, critical fugacity  $x_c$ is shown as a function of the interaction parameter  $u$. For $x\geq x_c$, that is on the critical line  and above it, the phase is polymerized with $\langle N\rangle=\infty$, whereas  below the critical line the phase is unpolymerized  with $\langle N \rangle=const$.  A polymerized phase on the critical line is  the expanded phase  for each value of   $u $. As one can see,  critical fugacity monotonically decreases  with the interaction parameter  $u$.  When $u\to \infty$ ($\epsilon\to-\infty$ or $T\to0$), it is found that    $x_c\to 0$ and SAWs  become   maximally compact. It is not difficult to see that  the growth constant of the maximally compact SAWs,  $\mu_C$, can be deduced from the critical fugacity in the $u\to\infty$ limit. For flexible SAWs, $\mathcal{G}(x,u)=\sum_Nx^NZ_N(u)$ where $Z_N(u)=\sum_MC_N(M)u^M$ is  canonical  partition function.  When $u\to\infty$, $Z_N$  is dominated by the term with the maximal number of contacts, that is,  $Z_N\approx u^{M_{max}}C_N({M_{max}})$  with $M_{max}=\frac{1}{2}(q-2)N$  being the  maximal number of contacts on a lattice with the coordination number $q$.  If we assume that the number of maximally compact SAWs,  $C_N(M_{max})$,  asymptotically grows as  $(\mu_C)^N$,  then    the radius of convergency of $\mathcal{G}$ is given by $x_c=\bigg(u^{\frac{q-2}{2}}\,\mu_C\bigg)^{-1}$. For 3-simplex lattice $q=3$ so that $\mu_C=\bigg(\sqrt{u}\,x_c\bigg)^{-1}$. Here we find $\mu_C\approx0.998$   for $u=50$  and    $x_c=0.14111...$, for example.    For larger $u$, calculated  values of $\mu_C$  tend toward  $1$, which is the exact value of the growth constant of Hamiltonian walks (HWs)  on the 3-simplex lattice~\cite{Sun}. Hamiltonian walks  have the same  'bulk' properties as the  maximally compact SAWs. However, there   might be  some   subtle differences regarding the  surface effects on  regular lattices that  are  still a  matter of debate~\cite{Prel, Owc0, Dup, Bai, Gut1}. On  fractal lattices, this  subject  is   much more complicated~\cite{Dus1,Dus2}.

  \begin{figure}
\includegraphics[width=0.465\columnwidth]{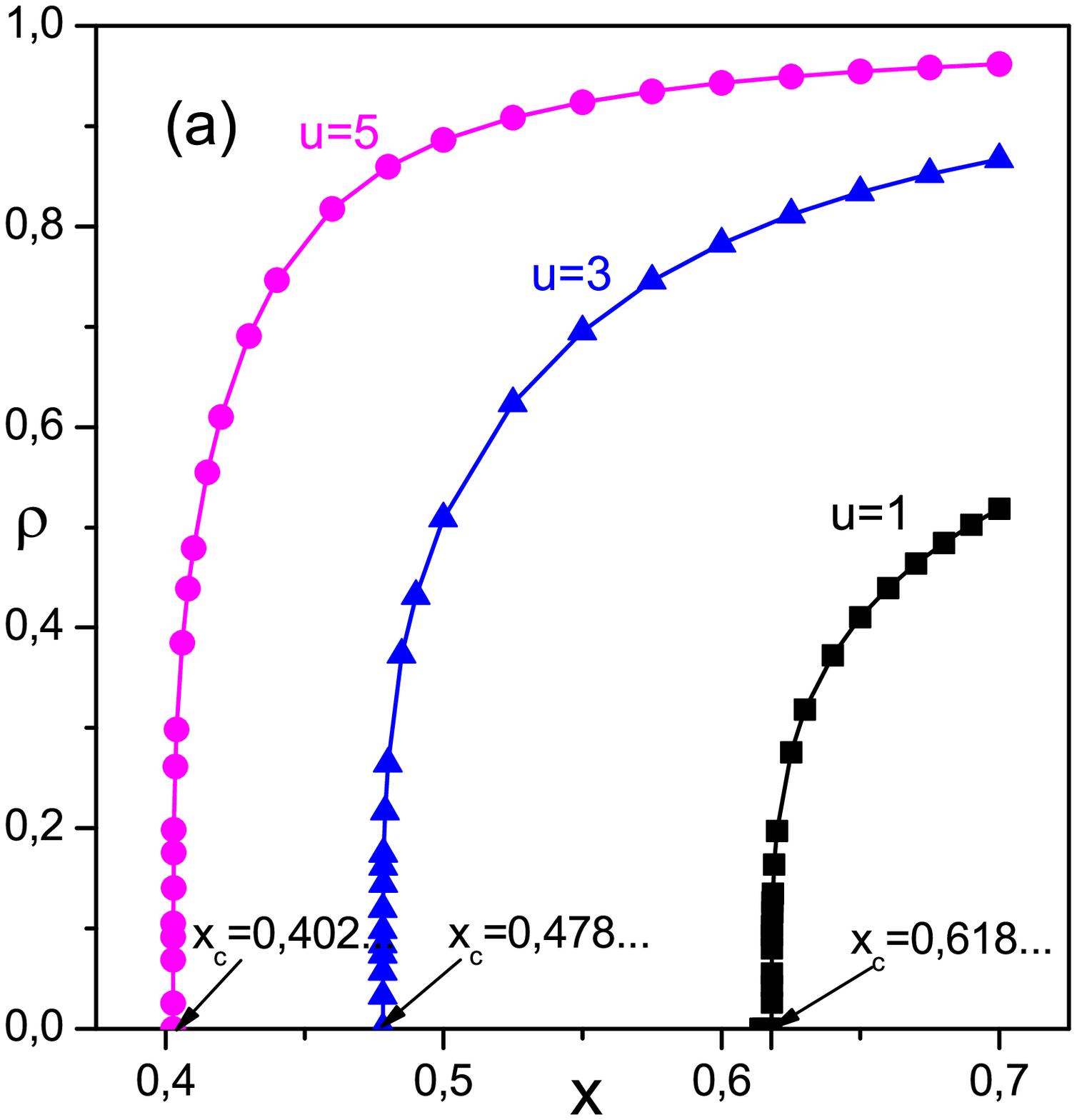}
\includegraphics[width=0.465\columnwidth]{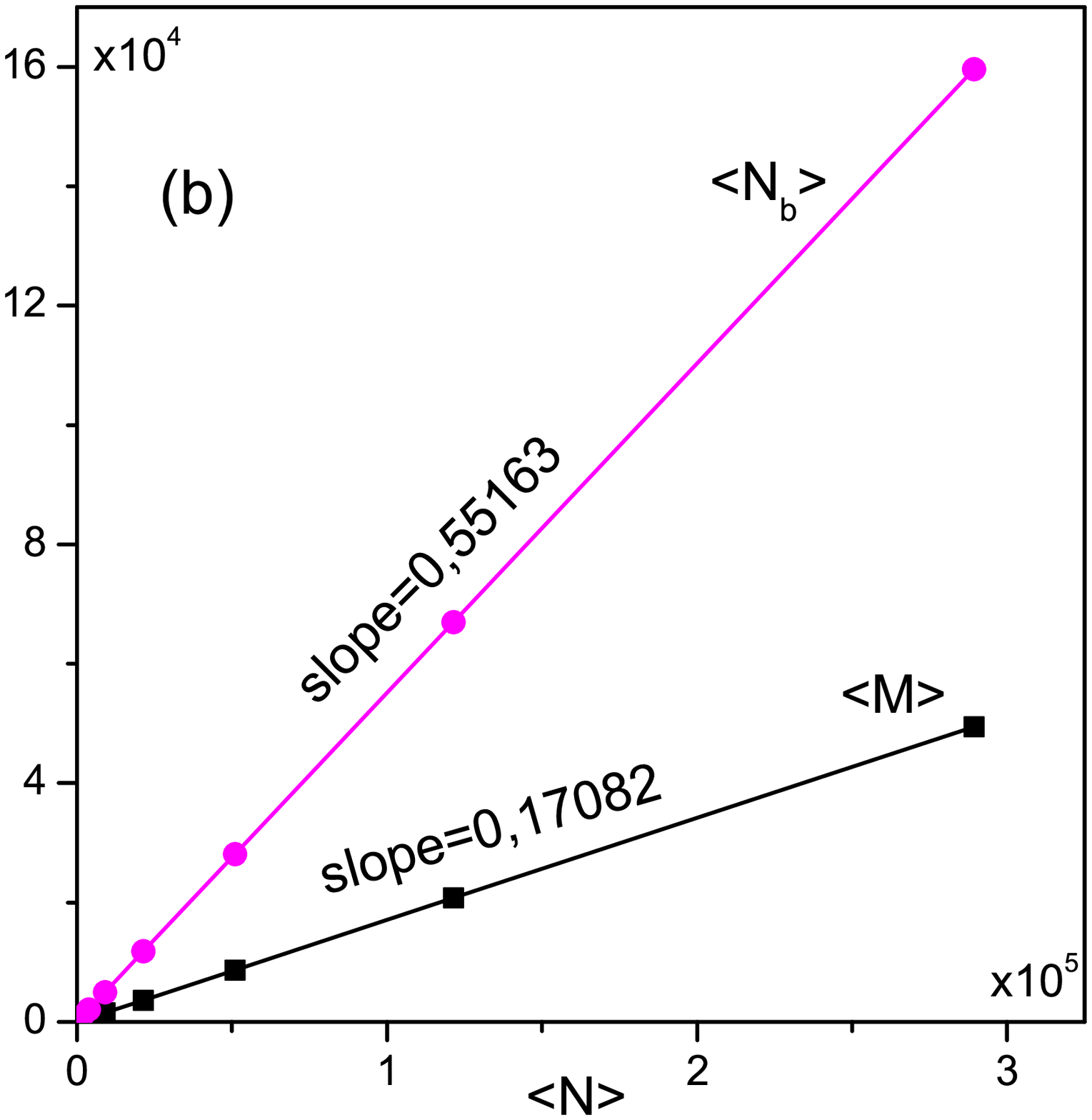}
\caption{ \label{fig:roMN}(a) Density of monomers  as a function of  fugacity $x$ for  fixed  values of $u$. (b) Asymptotic proportionality of the  mean number of contacts and  bends   to the mean number of  steps for $u=1$ and $s=1$. The slopes  $m$ and $n_b$ of the lines  are shown with  each graph. }
\end{figure}

\begin{figure}[t]
\begin{center}
\includegraphics[width=0.45\columnwidth]{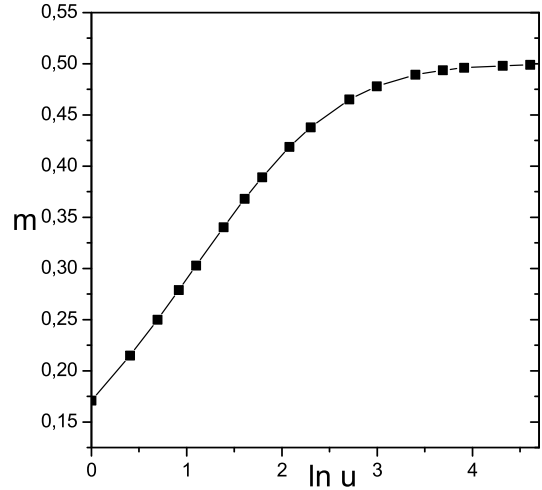}
\end{center}
\vspace{-0.5cm}
\caption{Mean number of contacts per mean number of steps  as a function of $ln(u)$.  }
\label{fig:mlnu}
\end{figure}

 The average density of monomers is  defined  as a fraction of the  $r$-th order generator's  sites  occupied by the  polymer,   $\rho_r=\langle N_r \rangle /3^r$.    This quantity is nonzero  in the polymerized phase above the  critical line (hence it is called   'dense' as marked in \Fref{fig:fugx}), whereas it is zero in the unpolymerized phase ('empty').  Density disappears continuously by crossing the critical line  from above at  $u<u_{\theta}$,  as  found  on regular \cite{Fost} and  fractal lattice \cite{Knez2} when the critical line is crossed above the $\theta$ temperature.   In  \Fref{fig:roMN}(a),  we show the  dependence of density on  fugacity for   three  given  values of $u$. One can see how density continuously vanishes at the critical fugacity   for every  $u$. When  $x\to \infty$ then    $\rho\to 1$,  which means that all sites of the lattice are occupied,  that is, SAWs become    Hamiltonian walks.  This limit is approached faster for larger values of $u$, as can be seen in  \Fref{fig:roMN}(a).

We now turn to the  mean number of contacts and bends. We have established linear relationships $\langle M \rangle\sim m \langle N \rangle$ and $\langle N_b\rangle\sim n_b\langle N\rangle$  when $N\to\infty$,  which is displayed in \Fref{fig:roMN}(b) for $u=1$ and $s=1$.  The same asymptotic behaviour   of the mean number of contacts is also  found on regular lattices \cite{Fis,Ish2,Ish3,Doug}. Asymptotic ratio $m$  is shown in \Fref{fig:mlnu}  as a function of $\ln u$. For smaller $u$ this dependence is almost linear (so that $m$ is a linear function of  $\epsilon$), similarly as on regular lattices   \cite{Ish3, Nem}. As  $u$ increases towards $\infty$,    $m$    asymptotically reaches a  maximal value  $m_{max}=(q-2)/2= 0.5$, which  it can have  on  the 3-simplex lattice.

\subsection{  Results and discussion, $0<s\leq1$}\label{rez2}

\begin{figure}
\includegraphics[width=0.52\columnwidth]{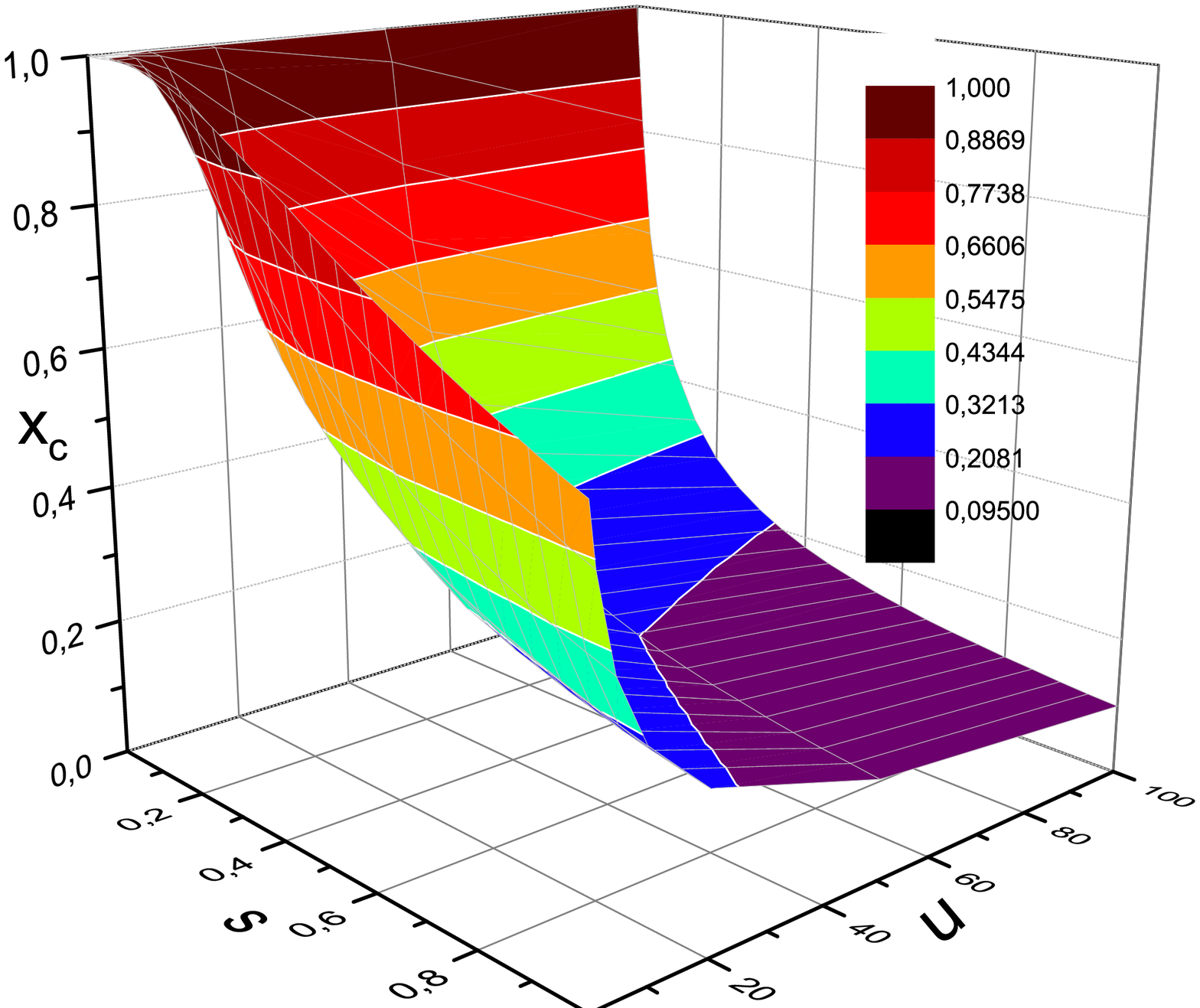}
\includegraphics[width=0.43\columnwidth]{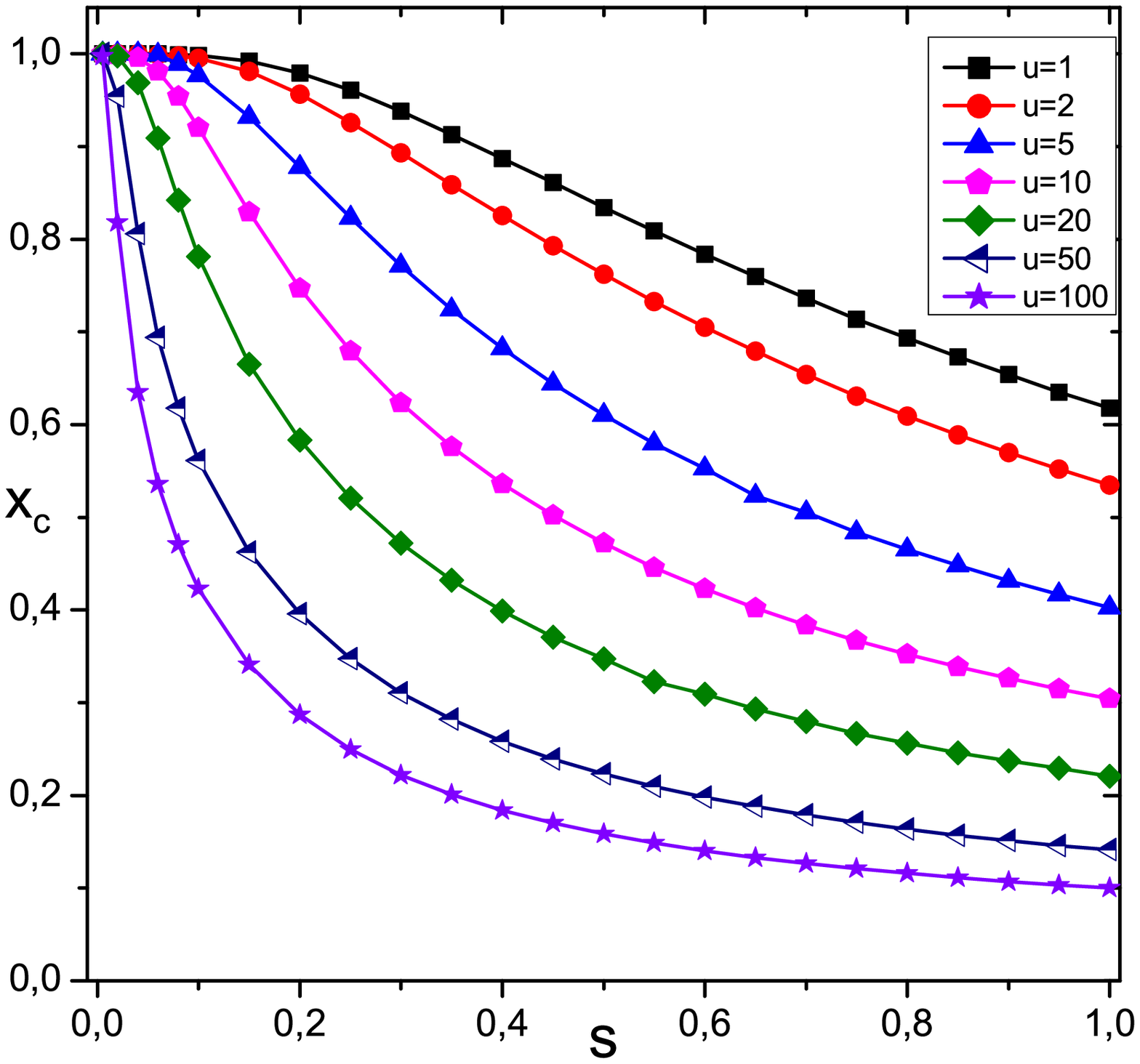}
\caption{ \label{fig:xc3d}Critical fugacity surface (left panel) and constant $u$  cross-sections (right panel).}
\end{figure}

 In this subsection we generalize results  obtained for  $x_c$, $m$ and $l_p$ on the case of semi-flexible interacting  walks. By varying $s$ over the  interval $(0,1]$ we obtain a critical surface  shown in    \Fref{fig:xc3d} (left panel).  For  each $s$ and $u$ on the critical surface polymer is in  the  expanded phase. Dependance of $x_c$  on $s$ for fixed values of $u$ is   shown on the right-hand  side panel.   For fixed $u$, $x_c$ is a  monotonically decreasing function of stiffness parameter (increasing function of stiffness energy). As $s$ decreases from $1$, $x_c$ increases and  tends toward  $ 1$   when $s\to 0$, for each  $u$.
\begin{figure}
\includegraphics[width=0.53\columnwidth]{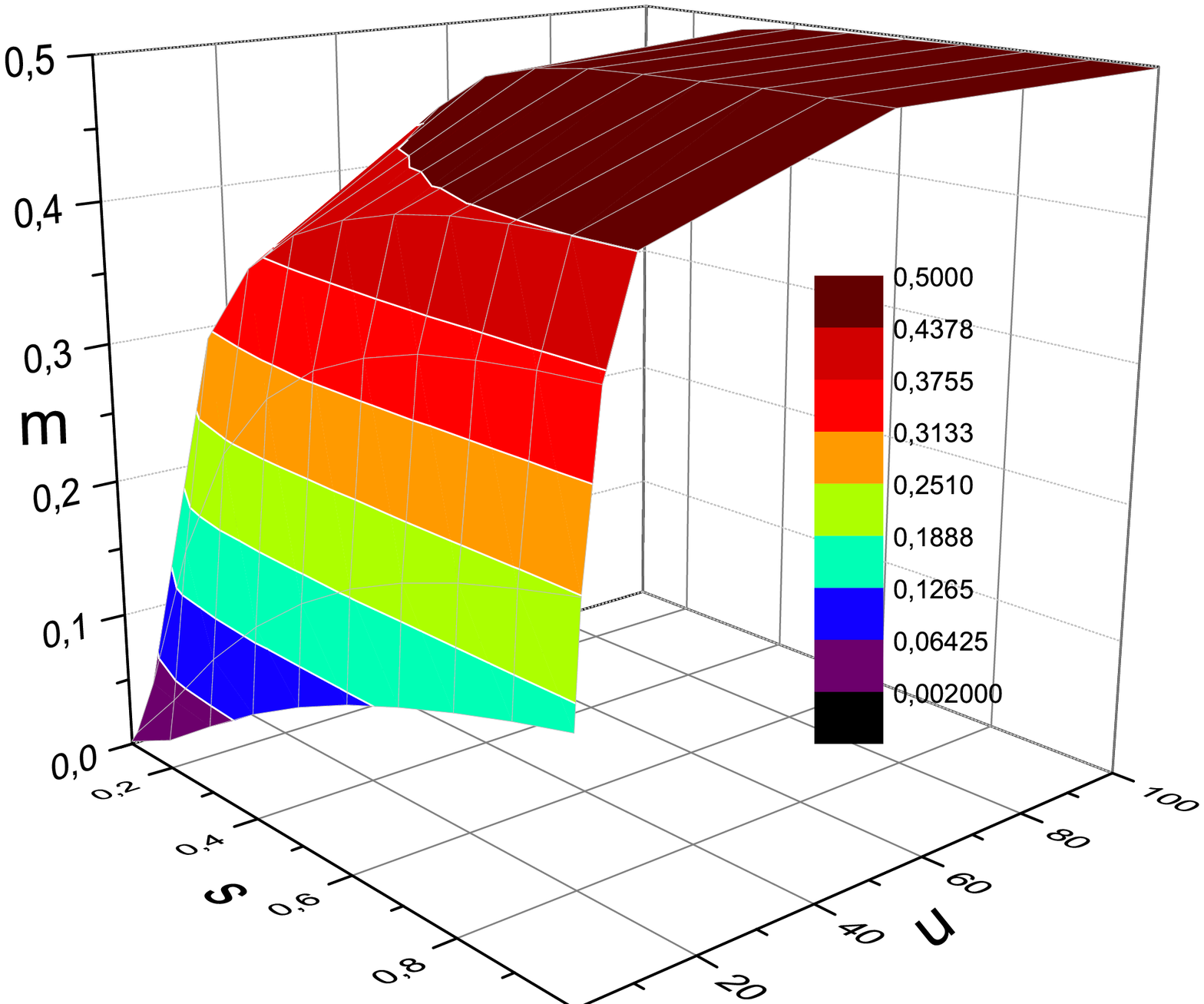}
\includegraphics[width=0.42\columnwidth]{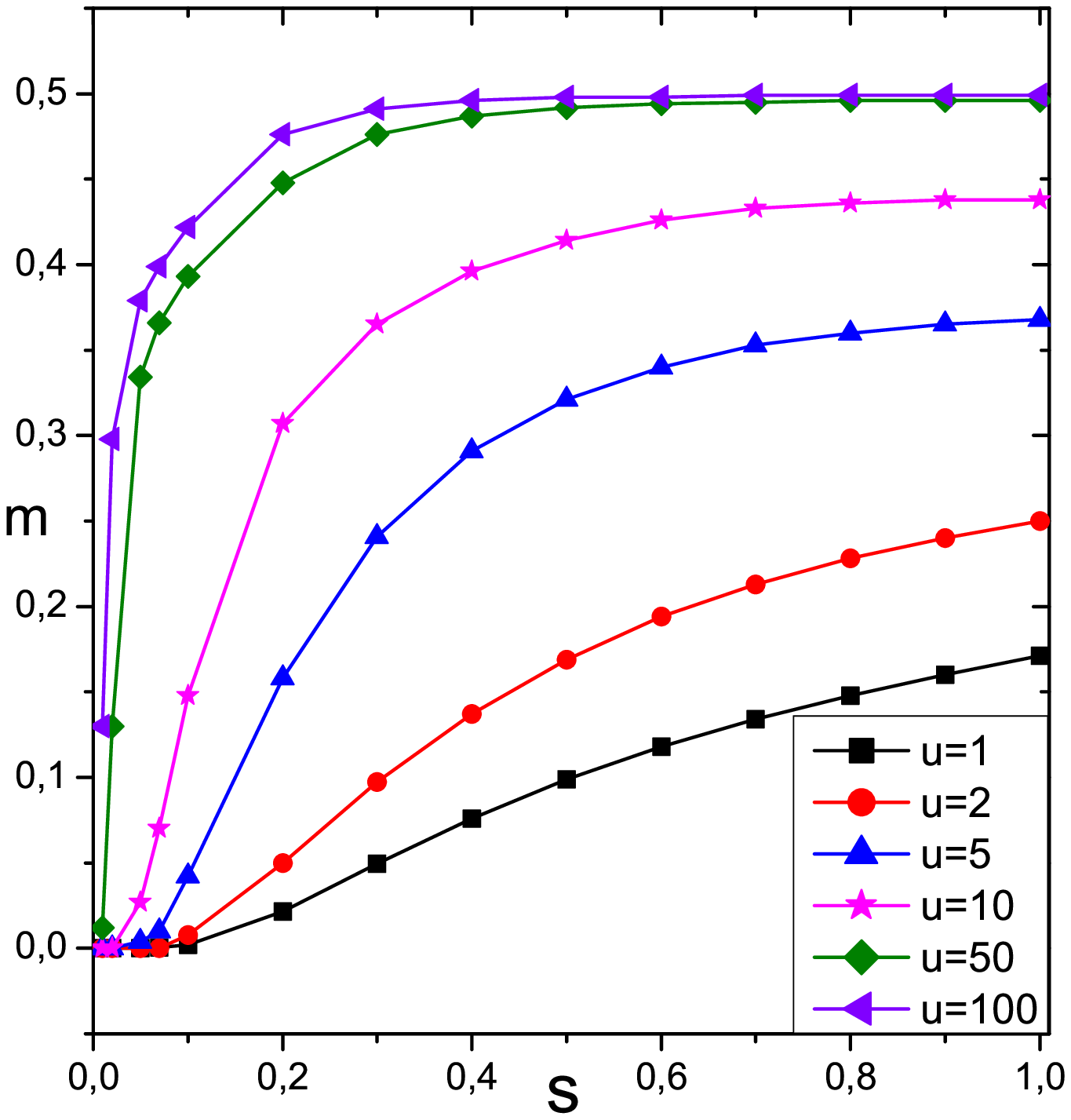}
\caption{Mean number of contacts  per mean number of steps as a  function of parameters $u$ and $s$ (left panel). Constant $u$ cross-sections (right panel).}
\label{fig:m}
\end{figure}

\begin{figure}[t]
\begin{center}
\includegraphics[width=0.5\columnwidth]{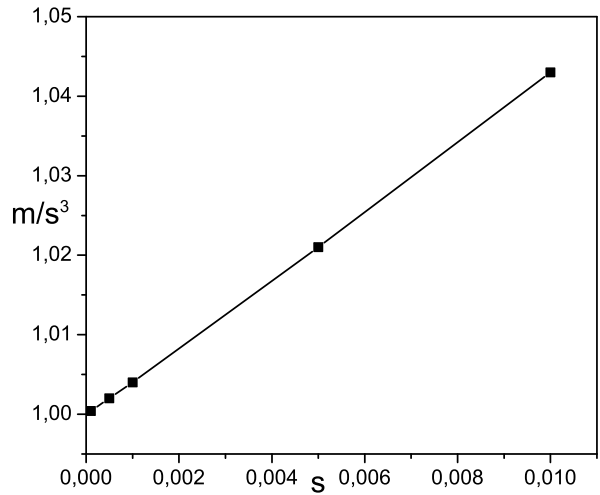}
\end{center}
\vspace{-0.4cm}
\caption{Ratio $m/s^3$ against $s$ for $10^{-4}\leq s\leq10^{-2}$.  }
\label{fig:mst0}
\end{figure}

\begin{figure}
\includegraphics[width=0.475\columnwidth]{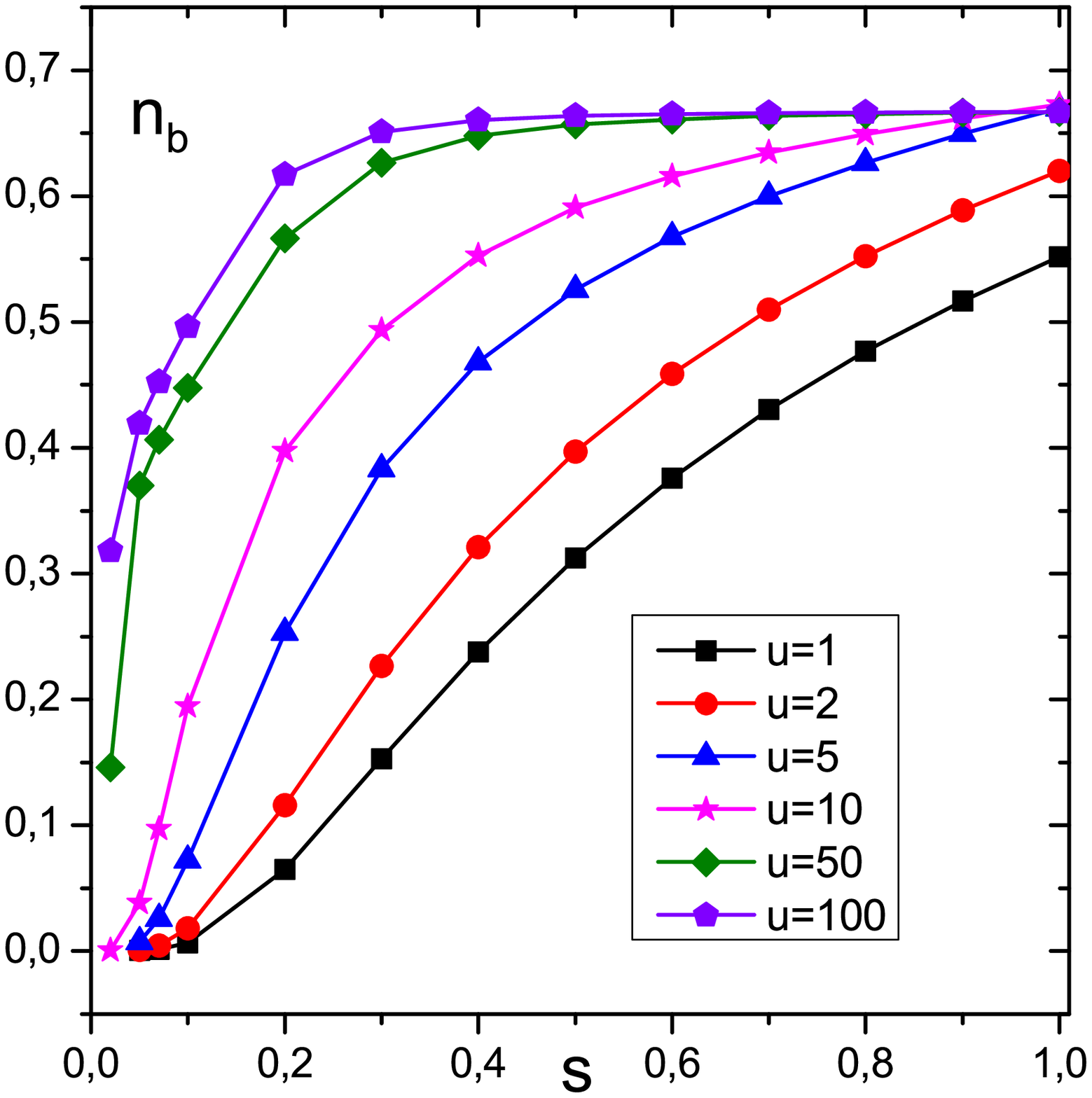}
\includegraphics[width=0.475\columnwidth]{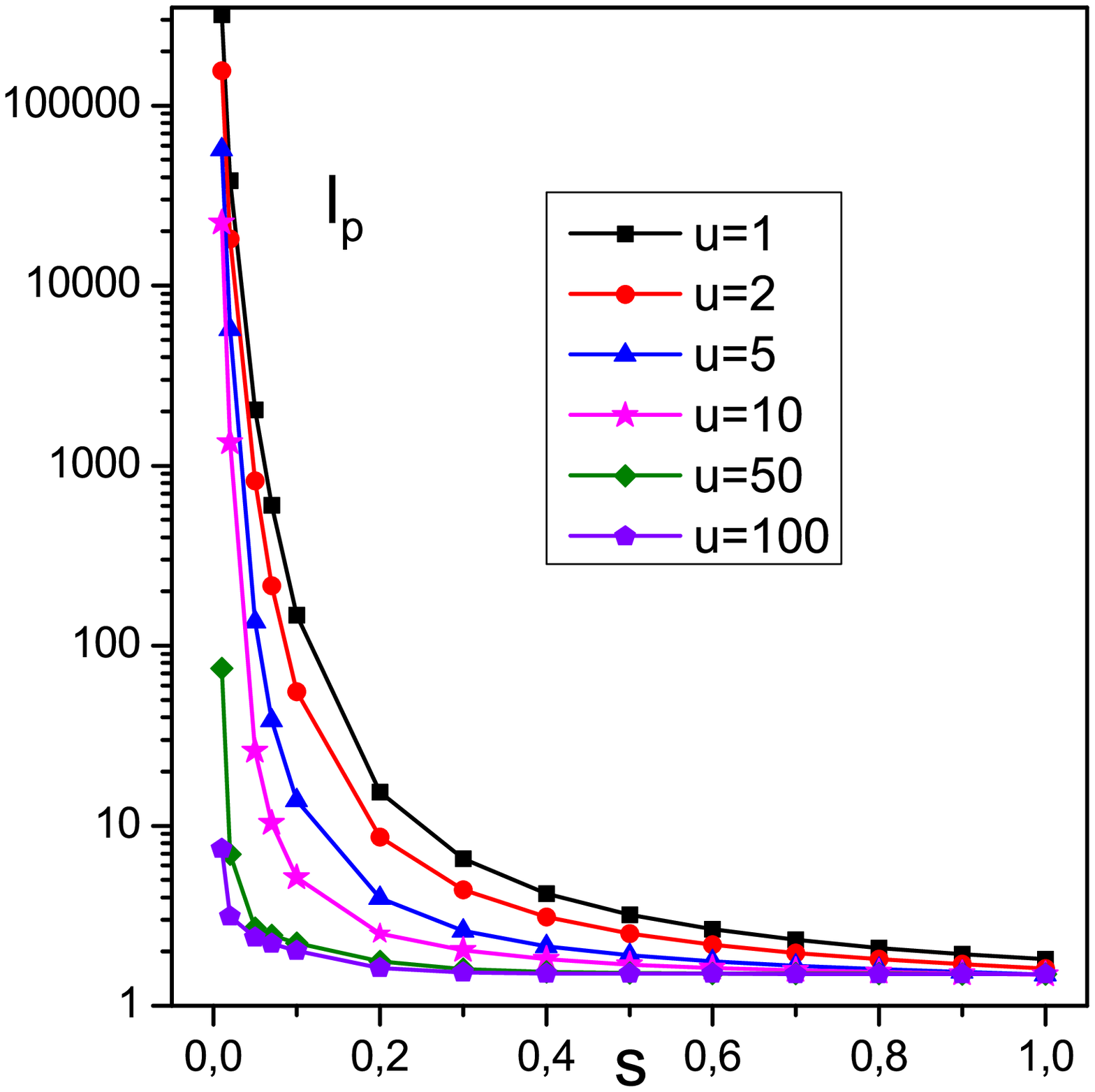}
\caption{Mean number of bends   per mean number of steps,  $n_b$, as a  function of   $s$  for fixed values of $u$ (left panel).   Persistence length $l_p=1/n_b$  as a function of $s$ for fixed values of $u$ (right panel).}
\label{fig:lp}
\end{figure}

\par   Linear relationships  $\langle M\rangle\sim m\langle N \rangle$  and  $\langle N_b\rangle\sim n_b\langle N \rangle$   in the limit  $N\to\infty$ are established  for all values of  $s$.  In  \Fref{fig:m}, $m(s,u)$ is presented as a surface in three-dimensional space (left panel) and $m(s)$ for fixed values of  $u$ (right panel). One can see that  $m$ is a monotonically  increasing function of interaction energy and   decreasing function of stiffness energy. When $u\to\infty$ ($\epsilon\to-\infty$) semi-flexible walks  become maximally compact as do flexible, although  for smaller $s$  (stiffer walks) this limit is approached much slower.   But, when $s\to 0$, SAWs tend to a rigid-rod limit with $m=0$ irrespectively of  $u$.  Analysing our numerical results we have found  that  $m\approx us^3$ for small $s$.  When $u=1$  this relation  starts to hold  for     $s\leq0.01$, and~\Fref{fig:mst0}  displays   how   limit   $m/s^3\to 1$ is approached when  $s$  is decreased  over two orders of magnitude.  Validity of  relation has been checked numerically for $s$  as small as $10^{-5}$.

\par     In \Fref{fig:lp} (left panel), $n_b$ is  displayed as a  function of $s$ for fixed  $u$, where it can be  seen that the   mean number of bends per mean number of steps  monotonically decreases with stiffness energy. In the right panel of \Fref{fig:lp}, $l_p(s)$  for constant $u$ is shown on  the logarithmic scale in order to capture  its  fast growth  for  $s<0.1$. One can see that the  interaction delays  this growth, but eventually all curves will terminate at  infinity as $s\to 0$.  We have found that the  persistence length diverges as  $l_p\sim1/(3us^3)$ when $s\to 0$. However, much faster divergency, although of differently defined persistence length,  was found in~\cite{Gia} on the same lattice.  There,  by applying  scaling analysis  on  semi-flexible SAWs, it is   concluded  that persistence length  should diverge as an exponential function of $1/s$ when $s\to 0$.   In our numerical approach we first calculate  critical fugacity $x_c$  (exact $x_c$  is known  only for  $s=1$ and  $u=1$), so  we deal with a  value  which is always slightly less than exact. We can  achieve very  high accuracy, for example,  $x_c=0.99999896840479804...$ is calculated with more than  seventy significant figures for $s=0.01$,  and  we can perform very large number of iterations, which is all quite straightforward  with a software like  Maple.  In such a way, we have  found that although variables show crossover behaviour with  the iterations, calculated quantities defined as ratios, do not. The  flow of variables  is exactly as was reasoned  in\cite{Gia}.  Variable $A_1$ (which contains  walks  composed of straight  steps)  stays close to its  initial value $x_c^2$ (just below $1$)  for the   first hundred  iterations, after which  it moves  toward fixed point   $A^*$,  where it stays nearby for almost next  hundred iterations.  Then it   iterates to zero (a value  that it has at the trivial attractive fixed point that governs the  empty phase). Other variables, starting  with their initial values, follow the similar trajectories.  This complicated behaviour is reflected on the  flow of the  mean number of steps and bends (as well as exponent $\nu$),  but not on   the ratio    $\langle N\rangle/\langle N_b\rangle$,  which  shows simple and fast convergency with the iterations.  Thus, we think that  persistence length considered in~\cite{Gia} and  our $l_p$ are different measures of persistency.

\par  Some of the  calculated values  for   $m$ and $l_p$ are presented in~\Tref{tab:m} and~\Tref{tab:lp}, respectively.    From~\Tref{tab:m}  we see that $m\approx 0.17082$ for ordinary SAWs, and  we would like to compare it  with the known values on regular  lattices, since, to the best of our knowledge, there are no  reported results on fractal ones. On  hypercubic lattices   with the  coordination number $q=2d$ in $d$-dimensional space   $m\approx0.1592(8),0.201(1)$ and  $0.174(2)$  for  $d=2,3$ and $4$, respectively,  thus largest in $d=3$, as  shown in \cite{Doug}. Somewhat later   result  $0.17088(5)$  for  $d=4$  is reported in \cite{Owc}. Beside dimension,   on regular lattices $m$ strongly depends on the lattice structure. In  the same  dimension  $m$ is larger on a  lattice with  larger coordination number, but it  diminishes with dimension  for the same coordination number. For the later example,    $m$ is much smaller  on the tetrahedral  lattice ($d=3$, $q=4$)  than on  the square lattice ($d=2$, $q=4$) \cite{Ish2}. However,  minimal number of steps needed to make a contact  ('minimum contact length'~ \cite {Doug}), on the tetrahedral (diamond) lattice is large, it is  5 in comparison with 3 on the square lattice.   One can expect that the  interplay between lattice  dimension and  other  properties of  fractal lattices   should influence $m$  even in a more complicated manner.  Coordination number of the $3$-simplex lattice is  3, its fractal dimension is $d_f\approx1.58$ and embedding dimension is  $d=2$, but $m$  exceeds the value  on  the square lattice.  Again, the   minimal contact length is just   2 which  supports  a larger $m$ obtained.  Regarding persistence length,  it  can be seen in~\Tref{tab:lp}  that $l_p$ is slightly a non-monotonic  function of  $u$ for $s=1$.    It is approximately $1.8128$ for  $u=1$ and  tends toward   $\frac{3}{2}$ from below when  $u\to\infty$.   For semi-flexible SAWs,  $l_p$  also attains the limit $\frac{3}{2}$ when  $u\to\infty$, albeit slower.  In this limit SAWs   become  maximally compact and have the same  $l_p$ as  Hamiltonian walks  on  the 3-simplex lattice~\cite{Dus3}.  Persistence length of semi-flexible  HWs is  independent  of stiffness and equal to  $\frac{3}{2}$  on 3-simplex lattice. On other fractal lattices, although stiffness dependent, $l_p$ of  semi-flexible HWs is  finite  even in the infinite stiffness limit~\cite{Dus3, Dus4}. Since Hamiltonian walks  visit all lattice sites and are enumerated on larger and larger lattices with  boundaries always present, they are    bounded by definition, which may reflect on $l_p$.  Finally, one last note regarding persistence length. Values  obtained here in the  non-interacting case ($u=1$)  are larger than the  values  of  equally defined $l_p$ for semi-flexible SAWs on the square lattice~\cite{Ziv}, and   for smaller $s$ this  discrepancy enlarges. This observation is  in  accord with the  expectation  that smaller space dimension should induce larger persistence length.
\par  In the  end, we would like to mention that  the mean number of contacts per mean number of steps and persistence length are quantities that  were suggested in~\cite{Bas}   as the  order parameters to  monitor  the  freezing transition  that the ISFSAW model  displayed on  the simple cubic lattice.   It  proved out  here  that these quantities together   can   reveal  the  local structure  of the  SAWs  conformations  characteristic for   the expanded phase found  for all  nonzero temperatures.  These conformations are  coil-like  with  'loops' of all sizes.  To illustrate this point and quantify the loops, we  recall  that  for ordinary SAWs   $m\approx0.1708$,  so that there are approximately  6 steps per contact, which is the size of a loop.  Also   $n_b\approx0.5516$, which means that  there are about  3 bends in the loop. For large $u$  loops are smaller and wigglier in the quasi zig-zag  conformations.   For small $s$  loops  are large,  then   both  $m$ and $n_b$ tend toward zero,  whereas  $n_b/m\approx3$ (for $s\neq 0$), which merely reflects  a geometrical constraint of the 3-simplex lattice.

\begin{table}
\caption{\label{tab:m} Some  numerical values   for $m$,   chosen from  a large set of data calculated and  used to produce graphs in~\Fref{fig:m}.   Five significant figures are presented with  the last digit rounded off.}

\begin{indented}
\lineup
\item[]\begin{tabular}{@{}*{6}{l}}
\br
 $ m $&$u=1$ &$u=2$&$u=5$&$u=10$&$u=100$\cr
\mr
$s=1$&0.170\,82  &0.249\,98&0.367\,96& 0.438\,30 &0.498\,91\cr
$s=0.8$&0.148\,16&0.228\,05& 0.360\,44& 0.436\,41& 0.498\,83\cr
$s=0.6$ & 0.117\,93&0.193\,69& 0.340\,34& 0.425\,75   &0.498\,32  \cr
$s=0.4$&0.075\,89 &0.137\,39& 0.290\,57&0.395\,78 &0.495\,92   \cr
$s=0.2$& 0.021\,356&0.049\,914&0.157\,86&0.307\,44&0.475\,55\cr
$s=0.1$& 0.002\,251\,7&0.007\,610\,4&0.042\,66&0.391\,68&0.421\,75\cr
$s=0.05$&0.000\,164\,00& 0.000\,464\,79&0.003\,9838&0.026\,91& 0.378\,53 \cr
\br
\end{tabular}
\end{indented}
\end{table}

\begin{table}
\caption{\label{tab:lp} Values of $l_p$ for some chosen $u$ and $s$. Five significant figures are presented with the last digit rounded off. }

\begin{indented}
\lineup
\item[]\begin{tabular}{@{}*{6}{l}}
\br
 $ l_p $&$u=1$ &$u=2$&$u=5$&$u=10$&$u=100$\cr
\mr
$s=1$&\0\0\01.8128  &\0\01.6133&\0\01.4927& \01.4856 &1.4999\cr
$s=0.8$&\0\0\02.0977&\0\01.8109&\0\01.5961&\01.5403& 1.5012\cr
$s=0.6$ & \0\0\02.6590&\0\02.1789&\0\01.7619& \01.6236   &1.5037  \cr
$s=0.4$&\0\0\04.2033 &\0\03.1162&\0\02.1359&\01.8086 &1.5139  \cr
$s=0.2$&\0\015.381&\0\08.6237&\0\03.9493&\02.5153&1.6214\cr
$s=0.1$&\0148.06&\055.551&\013.841&\02.2337&2.0162\cr
$s=0.05$&2\,034.8& 823.82&135.44&26.020& 2.3861 \cr
\br
\end{tabular}
\end{indented}
\end{table}

\section{ Summary and conclusions}\label{cetvrta}
In this paper we have studied interacting semi-flexible self-avoiding walks and polygons on the 3-simplex lattice,  which can serve as  a model of linear and ring polymer conformations  in a nonhomogeneous environment. Analysis of the closed set of   established recurrence equations has    shown that  for all non-zero values of  the stiffness parameter $s$  and finite  values of the interaction parameter $u$, model describes a polymer in an expanded, swollen phase.  By applying  the  grand canonical formalism, non-universal properties such us  the mean number of contacts and the  mean number of bends  are calculated for the   set of values satisfying $0\leq u<\infty$ and $0< s\leq1$.  Linear relationships $\langle M \rangle\sim m\langle N\rangle$ and $\langle N_b \rangle\sim n_b\langle N\rangle$    are  established in the limit $N\to\infty$,  and the  ratios    $m$ and $n_b$  are  calculated numerically for various values of $u$ and $s$. It is perceived that the  calculated quantities  should be valid  equally well for self-avoiding walks and polygons since  the  critical fugacity which determines free energy in the thermodynamic limit is the same for both types of walks.    Results for $m$ and $l_p=1/n_b$  (which is a  measure  of  polymer persistency)  are presented and  comprehensively  discussed  in subsections~\ref{rez1} and~\ref{rez2}.   Disorder modeled by the 3-simplex lattice brought about larger values of   $m$ and $l_p$   obtained for ordinary   SAWs  in comparison  with regular lattices.  Interaction  and stiffness energies  influenced  $m$ and $l_p$   independently, which resulted in  monotonic behaviour. An exception   is a  slight non-monotonicity of $l_p$ as a function of $u$ for $s=1$.  Limiting  values of  parameters have also been considered. Semi-flexible   SAWs  in  the limit $u\to \infty$ become maximally compact, with    $m$ and $l_p$  that coincide  with the results for  Hamiltonian walks on  the 3-simplex lattice.   However, in the limit of  infinite stiffness ($s\to0$), stiffness  completely prevails and conformations with the infinite persistence length, with no contacts,  at  $s=0$ remain.  We have  found that in this limit the mean number of contacts per mean number of steps tends toward zero as    $m\approx u s^3$, while the persistence length diverges as   $l_p\sim1/(3us^3)$ .  At exactly  $s=0$ polymer is a rigid-rod.  Since there have been no  other results on this subject, it would  be worthwhile to  conduct a  similar research on other  fractal lattices to learn more about   the influence of fractal space  and interactions on these   nonuniversal  properties of SAWs.

\appendix
\section{}

\begin{figure}[t]
\begin{center}
\includegraphics[scale=0.8]{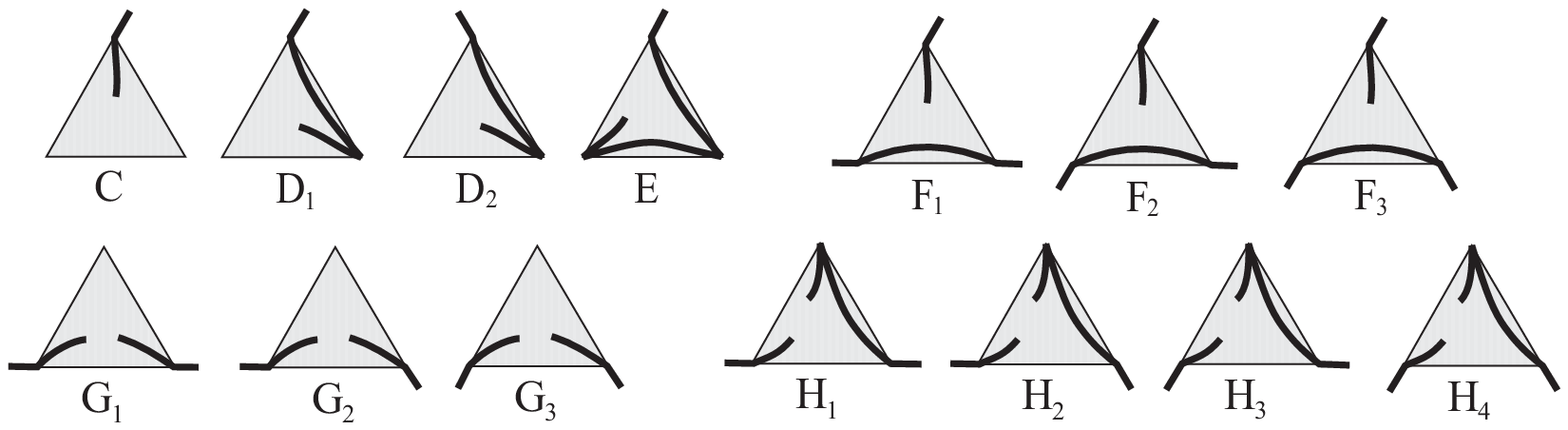}
\end{center}
\vspace{-0.3cm}
\caption{ Walks with one or both endpoints in the interior vertices  of the  generator classified according to the number of visited corner vertices  and direction of the  first steps outside of  the generator.  }
\label{fig:CD}
\end{figure}
  We supply additional types of walks  that are involved in   recursive formulation of the generating function (\ref{eq:genG}).  These  are   walks with one or both endpoints in the interior of the generator and   are schematically presented in  \Fref{fig:CD}.   $C$ is the weight of all SAWs that starts in the interior of the generator and leave it through one of the corner vertices with the external step along the fixed triangle side.  These walks are symmetrically  related with the walks  directed along the other triangle side and thus have the same weight. They include walks that visit none, one  and both of the remained corner vertices. Other walks are similarly defined.    Some of them are supsets of others: $C\supset D_1\supset E$, $G_1\supset H_1$, $G_2\supset H_2$, $G_3\supset H_4$. Their   recurrence equations are given by:

\begin{eqnarray}
  C'&=&C+(A_1+A_2)C+(A_1+A_2)A_3C+(F_1+F_2)A_3^2+\nonumber\\
  &+& (u-1)(B_1+B_2)A_3D_1\,,
\end{eqnarray}
\begin{eqnarray}
  D_1' &=& A_2B_3C+(A_1A_3+A_2)D_2+(F_1+F_2)A_3B_3+ \nonumber \\
   &+& (u-1)(B_2B_3D_1+A_3B_1E) \,,
   \end{eqnarray}
   \begin{eqnarray}
    D_2' &=& A_1B_3C+(A_2A_3+A_1)D_2+(F_1+F_2)A_3B_3+ \nonumber\\
   &+& (u-1)(B_1B_3D_1+A_3B_2E)\,,
\end{eqnarray}
\begin{equation}
E' =(A_1+A_2)B_3D_2+(F_1+F_2)B_3^2+  (u-1)(B_1+B_2)B_3E\,,
\end{equation}
\begin{eqnarray}
  F_1' &=& A_1^2C+ A_2^2F_3+A_1^2F_1+A_1A_2F_1+(u-1)\Big[(D_1+D_2)A_1B_1 \nonumber \\
   &+& (B_1+B_2)B_1F_1+B_2^2F_3+(u-1)B_1^2E\Big] \,,
  \end{eqnarray}
  \begin{eqnarray}
    F_2' &=&A_1A_2C+ A_1A_2F_3+A_1^2F_2+A_2^2F_1+(u-1)\Big[A_1B_2D_2 \nonumber \\
   &+& A_2B_1D_1+B_1B_2F_3+B_1^2F_2+B_2^2F_1+(u-1)B_1B_2E\Big] \,,
    \end{eqnarray}
     \begin{eqnarray}
    F_3' &=& A_2^2C+ A_1^2F_3+A_1A_2F_2+A_2^2F_2+ (u-1)\Big[(D_1+D_2)A_2B_2\nonumber \\
   &+&(B_1F_3+B_2F_2)B_1+B_2^2F_2+(u-1)B_2^2E\Big] \,,
\end{eqnarray}
\begin{eqnarray}
  G_1' &=&C^2+2(F_1F_2+F_1F_3+F_2F_3)A_3+2A_2C^2+  2(F_2+F_3)A_1C\nonumber \\
   &+&(2A_3G_2+A_2G_3)A_2+2A_1G_1+(u-1)\Big[D_2^2+2A_2D_1^2\nonumber \\
   &+& 2B_2CD_2+2(F_2+F_3)B_1D_1 +2(H_2+H_3)A_3B_2+B_2^2G_3+\nonumber\\
   &+&2(u-1)B_2D_1E\Big]\,,
\end{eqnarray}
\begin{eqnarray}
  G_2' &=&C^2+2(F_1+F_2)A_3F_3+ (F_1^2+F_2^2)A_3+(A_1+A_2)C^2 \nonumber\\
  &+& (A_1+A_2)CF_3+(A_1F_1+A_2F_2)C+(A_1G_2+A_2G_1)A_3\nonumber \\
   &+&(A_2G_3+G_2)A_1+A_2G_1+(u-1)\Big[A_1D_1^2+(1+A_2)D_1D_2 \nonumber \\
   &+&(B_2D_1+B_1D_2)C+(B_1+B_2)D_1F_3+(B_1F_1+B_2F_2)D_1\nonumber\\
   &+&(H_2+H_3)A_3B_1+2A_3B_2H_1+B_1B_2G_3\nonumber\\
   &+& (u-1)(B_1+B_2)D_1E\Big]\,,
\end{eqnarray}
\begin{eqnarray}
  G_3' &=&C^2+2(F_1F_2+F_1F_3+F_2F_3)A_3+2A_1C^2+  2(F_1+F_3)A_2C\nonumber \\
   &+&(2A_3G_1+A_1G_3)A_1+2A_2G_2+(u-1)\Big[D_1^2+2A_1D_1D_2\nonumber \\
   &+& 2B_1CD_1+2(F_1+F_3)B_2D_1 +4A_3B_1H_1+B_1^2G_3+\nonumber\\
   &+&2(u-1)B_1D_1E\Big]\,,
\end{eqnarray}
\begin{eqnarray}
  H_1' &=&A_2^2H_4+(F_1F_2+F_1F_3+F_2F_3)B_3+A_2CD_2+  (F_2+F_3)A_1D_2\nonumber \\
   &+&A_2B_3G_2+(u-1)\Big[B_2^2H_4+A_2D_1E+ B_2D_2^2+(F_2+F_3)B_1E \nonumber\\
   &+&(H_2+H_3)B_2B_3+(u-1)B_2E^2\Big]\,,
\end{eqnarray}
\begin{eqnarray}
  H_2' &=&A_1A_2H_4+(F_1+F_2)B_3F_3+B_3F_1^2+A_1CD_2+  (A_1F_1+A_2F_3)D_2\nonumber \\
   &+&A_1B_3G_2+(u-1)\Big[B_1B_2H_4+A_1D_1E+ B_1D_2^2 \nonumber\\
   &+&(B_1F_1+B_2F_3)E+(H_2+H_3)B_1B_3+(u-1)B_1E^2\Big]\,,
\end{eqnarray}
\begin{eqnarray}
  H_3' &=&A_1A_2H_4+(F_1+F_2)B_3F_3+B_3F_2^2+A_2CD_2+  (A_1F_3+A_2F_2)D_2\nonumber \\
   &+&A_2B_3G_1+(u-1)\Big[B_1B_2H_4+A_2D_2E+ B_2D_1D_2 \nonumber\\
   &+&(B_1F_3+B_2F_2)E+2B_2B_3H_1+(u-1)B_2E^2\Big]\,,
\end{eqnarray}
\begin{eqnarray}
  H_4' &=&A_1^2H_4+(F_1F_2+F_1F_3+F_2F_3)B_3+A_1CD_2+  (F_1+F_3)A_2D_2\nonumber \\
   &+&A_1B_3G_1+(u-1)\Big[B_1^2H_4+A_1D_2E+ B_1D_1D_2 \nonumber\\
   &+&(F_1+F_3)B_2E+2B_1B_3H_1+(u-1)B_1E^2\Big]\,.
\end{eqnarray}
Initial conditions are: $C(1)=x^{1/2}(1+x+xs+x^2su+x^2s^2u)$, $D_1(1)=x^{3/2}s(1+xsu+xu)$, $D_2(1)=x^{3/2}(1+xsu+xs^2u)$, $E(1)=x^{5/2}su(1+s)$, $F_1(1)=x^{5/2}u^2$, $F_2(1)=x^{5/2}su^2$, $F_3(1)=x^{5/2}s^2u^2$, $G_1(1)=xu(1+2xsu)$, $G_2(1)=xu(1+xu+xsu)$, $G_3(1)=xu(1+2xu)$,  $H_1(1)=x^2su^2$, $H_2(1)=x^2u^2$, $H_3(1)=x^2su^2$ and $H_4(1)=x^2u^2$. On the unit triangle, fugacity $x$ is assigned to each vertex  through which the walk passes by  and fugacity $x^{1/2}$  to each vertex which is a starting (ending) point of the walk, in order that each step of the whole walk gets proper  fugacity $x$.


\section*{References} \begin{thebibliography}{60}




\bibitem{Mad}  Madras~N and  Slade G 1993 {\it The Self-Avoiding walk} (Boston: Birkhaüser) p~435

\bibitem{Gut}    Guttmann~A~J ed 2009 {\it Polygons, Polyominoes and Polycubes} (vol 775 Lecture notes in physics)  (Berlin:Springer) p~499
\bibitem{Ren} Janse van Rensburg~E~J 2015  {\it The Statistical Mechanics of Interacting Walks, Polygons, Animals and Vesicles} (Oxford Lecture Series in Mathematics and Its Applications)   (Oxford: Oxford University Press) p~625

\bibitem{Or} Orr~W~J~C  1947 Statistical treatment of polymer solutions at infinite dilution {\it Trans. Faraday Soc.} {\bf43} 12\textendash27
\bibitem{Flori} Flory~P~J 1949  The configuration of real polymer chains {\it J. Chem. Phys.} {\bf 17} 303\textendash10


     \bibitem {Ish1}  Ishinabe~T 1985 Examination of the theta-point from exact enumeration of self-avoiding walks  {\it J. Phys. A: Math. Gen.} {\bf 18}   3181\textendash87
      \bibitem{Fis} Fisher~M~E and Hiley~B~J 1961 Configuration and Free Energy of a Polymer Molecule with Solvent Interaction {\it J. Chem. Phys.} {\bf 34}    1253\textendash67
     \bibitem {Ish2}  Ishinabe~T and Chikahisa~Y  1986 Exact enumerations of self-avoiding lattice walks with different nearest-neighbor contacts  {\it J. Chem. Phys.} {\bf 85}   1009\textendash17
    \bibitem{Ish3}  Ishinabe~T 1987 Examination of the theta-point from exact enumeration of self-avoiding walks II {\it J. Phys. A: Math. Gen.} {\bf 20 }  6435\textendash53
    \bibitem{Doug}  Douglas~J~F and  Ishinabe~T 1995 Self-avoiding-walk contacts and random-walk self-intersections in variable dimensionality {\it Phys. Rev. E} {\bf 51}   1791\textendash817
        \bibitem{Der}  Derrida~B and  Saleur~H  1985  Collapse of two-dimensional linear polymers: a transfer matrix calculation of the exponent $\nu_t$ {\it J. Phys. A: Math. Gen.} {\bf 18}   1075\textendash79

\bibitem{Sal}  Saleur~H 1986  Collapse of two-dimensional linear polymers {\it J. Stat. Phys.}  {\bf 45}  419\textendash38

\bibitem{Bin} Binder~P~M, Owczarek~A~L,  Veal~A~R and Yeomans~J~M 1990 Collapse transition in a simple polymer model: exact results {\it J. Phys. A: Math. Gen} {\bf 23} L975\textendash79
\bibitem{Nem}  Nemirovsky~A~M,  Dudowicz~J and Freed~K~F 1992 Thermodynamics of a Dense Self-Avoiding Walk with Contact Interactions {\it J. Stat. Phys.}  {\bf 67}  395\textendash412
\bibitem{Gras}  Grassberger~P and Hegger~R 1995  Simulations of three-dimensional $\theta$ polymers {\it J. Chem. Phys.} {\bf 102} 6881\textendash89

\bibitem{Bar}  Barkema~G~T and  Flesia~S 1996 Two-Dimensional Oriented Self-Avoiding Walks with Parallel Contacts{\it J. Stat. Phys.} {\bf 85}    363\textendash81
 \bibitem{Nid}  Nidras~P~P 1996  Grand canonical simulations of the interacting self-avoiding walk model {\it J. Phys. A: Math. Gen.} {\bf 29} 7929\textendash42
\bibitem{Tesi2} Tesi~M~C,  Janse van Rensburg~E~J, Orlandini~E and  Whittington~S~G 1996 Interacting self-avoiding walks and polygons in three dimensions {\it J. Phys. A: Math. Gen. 29} 2451\textendash63
 \bibitem{Ben} Bennett-Wood~D, Enting~I~G,  Gaunt~D~S, Guttmann~A~J,  Leask~J~L,  Owczarek~A~L and   Whittington~S~G  1998  Exact enumeration study of free energies of interacting polygons and walks in two dimensions {\it J. Phys. A: Math. Gen.} {\bf 31}  4725\textendash41
     \bibitem{Fost}  Foster~D~P  and  Seno~F 2001  Two-dimensional self-avoiding walk with hydrogen-like bonding: phase diagram and critical behaviour {\it  J. Phys. A: Math. Gen.}{\bf 34} 9939\textendash57
 \bibitem{Vog}  Vogel~T, Bachmann~V and Janke~W 2007 Freezing and collapse of flexible polymers on regular lattices in three diensions {\it Phys. Rev. E} {\bf 76} 061803.

 \bibitem{Pon}  Ponmurugan~M and  Satyanarayana~S~V~M 2012 The $\theta$ points of interacting self-avoiding walks and rings on a 2D square lattice {\it J. Stat. Mech.}   P06010
 \bibitem{Beat} Beaton~N~R, Guttmann~A~J  and   Jensen~I 2020 Two-dimensional interacting self-avoiding walks: new estimates for critical temperatures and exponents {\it J. Phys. A: Math. Theor.} {\bf 53}   165002

  \bibitem{Kol}  Kolinski~A,  Skolnick~J and  Yaris~R 1986 The collapse transition of semiflexible polymers. A Monte Carlo simulation of a
model system {\it J.  Chem. Phys.}{\bf 85}  3585\textendash97
\bibitem{Bas}   Bastolla~U  and Grassberger~P 1997 Phase transitions of single semistiff polymer chains {\it J. Stat. Phys.} {\bf 89} 1061\textendash78
\bibitem{Doye} Doye~J~P~K,  Sear~R~P and  Frenkel~D 1998 The effect of chain stiffness on the phase behaviour of isolated homopolymers {\it J. Chem. Phys.} {\bf 108} 2134\textendash42
\bibitem{Lise} Lise~S, Maritan~A and  Pelizzola~A 1998 Bethe approximation for a semiflexible polymer chain {\it Phys. Rev. E} {\bf 58} R5241\textendash 44
\bibitem{Kraw} Krawczyk~A,  Owczarek~A~L and  Prellberg~T 2009 Semi-flexible hydrogen-bonded and non-hydrogen bonded lattice polymers {\it Physica A } {\bf 388} 104\textendash12
\bibitem{Chak}  Chakrabarti~B~K  ed 2005 {\it Statistics of Linear Polymers in Disordered Media}  (Amsterdam: Elsevier) p~369
\bibitem{Bred}  Bradly~C~J and   Owczarek~A~L 2021 Effect of Lattice Inhomogeneity on Collapsed Phases of Semi-stiff ISAW Polymers {\it J. Stat. Phys.} {\bf 182} 27


 \bibitem{Klein} Klein~ D~J and  Seitz~W~A 1984   Self-interacting self-avoiding walks on the Sierpinski gasket {\it J. Physique Lett.} {\bf 45 }   241\textendash47
\bibitem {Dhar1}  Dhar~D and Vannimenus~J 1987  The collapse transition of linear polymers on fractal lattices {\it J. Phys. A: Math. Gen.} {\bf 20 } 199\textendash213
\bibitem {Knez1}  Kne\v{z}evi\'{c}~M and  Vannimenus~J 1987 Topological frustration and quazicompact phase in a model of interacting polymers {\it  J. Phys. A: Math. Gen.} {\bf 20}  L969\textendash 73
\bibitem{Kumar} Kumar~S and Singh~Y 1990  Collapse transition of linear polymers on a family of truncated n-simplex lattices, {\it Phys. Rev. A}{\bf 42}  7151\textendash54
\bibitem {Knez2}  Kne\v{z}evi\'{c}~D,  Kne\v{z}evi\'{c}~M and Milo\v{s}evi\'{c}~S 1992 Critical behavior of an interacting polymer chain in a porous model system: Exact results for truncated simplex lattices 1992 {\it Phys. Rev. B} {\bf 45} 574\textendash85
    \bibitem {Ziv0}  \v{Z}ivi\'{c}~I,  Milo\v{s}evic\'{c}~S and Djordjevi\'{c}~B 2005 On the total number of distinct self-interacting
self-avoiding walks on three-dimensional fractal structures 2005 {\it J. Phys. A: Math. Gen.} {\bf 38}  555\textendash65
   \bibitem{Gia} Giacometti~A and  Maritan~A 1992  Self-avoiding walks with curvature energy on fractals {\it J. Phys. A: Math. Gen} {\bf 25} 2753\textendash64
\bibitem{Tut} Tuthill~G~F  and Schwalm~W~A 1992  Biased interacting self-avoiding walks on the four-simplex lattice {\it Phys.Rev. B} {\bf 46}  13722\textendash34

 \bibitem{Dar2}  Dhar~D 1978  Self-avoiding random walks: Some exactly soluble cases  {\it J. Math. Phys.} {\bf 19} 5\textendash11
\bibitem{Dhar3} Dhar~D 1977  Lattices of efectively nonintegral dimensionality {\it J. Math. Phys. }{\bf 18}  577\textendash85
\bibitem{Ram}  Rammal~R,  Toulouse~G and  Vannimenus~J 1984  Self-avoiding walks on fractal spaces: exact results and Flory approximation {\it Journal de Physique} {\bf 45}  389\textendash94
     \bibitem{Pol}  Polotsky~A~A and   Ivanova~A~S 2021  On the adsorption of a polymer chain with positive or negative bending stiffness onto a planar surface {\it Physica A} {\bf 562} 125319
  \bibitem{Ziv}   \v{Z}ivi\'{c}~I,  Had\v{z}i\'{c}~S~E and Mar\v{c}eti\'{c}~D 2022  Persistence length of semi-flexible polymer chains on Euclidean lattices {\it  Physica A} {\bf 607}  128222

\bibitem{Sun}  Elezović-Hadžić~S,   Marčetić~D and  Maletić~S 2007  Scaling of Hamiltonian walks on fractal lattices {\it Phys. Rev. E} {\bf 76}  011107


\bibitem{Owc0}  Owczarek~A~L, Prellberg~T and  Brak~R 1993 New scaling form for the collapsed polymer phase {\it Phys. Rev. Lett.} {\bf 70} 951\textendash53
\bibitem{Prel}   Prellberg~T,  Owczarek~A~L, Brak~R and  Guttmann~A~J 1993  Finite-length scaling of collapsing directed walks {\it Phys. Rev. E} {\bf 48}  2386\textendash96
\bibitem{Dup} Duplatier~B  1993  Exact Scaling Form for the Collapsed 2 D PolymerPhase {\it Phys. Rev. Lett.} {\bf 71} 4274
\bibitem{Bai}  Baiesi~M,  Orlandini~E and Stella~A~L 2005 Scaling of a Collapsed Polymer Globule in Two Dimensions {\it Phys. Rev. Lett.} {\bf 96 } 040602
\bibitem{Gut1}  Guttmann~A~J, Jensen~I and Owczarek~A~L 2022  Self-avoiding walks contained within a square {\it J. Phys. A: Math. Theor.} {\bf 55}  425201
\bibitem{Dus1} Lekić~D and  Elezović-Hadžić~S 2010 A model of compact polymers on a family of three-dimensional fractal lattices {\it J. Stat. Mech.}  P02021
\bibitem{Dus2} Marčetić~D,  Elezović-Hadžić~S  and \v{Z}ivi\'{c}~I 2021 Effects of the boundaries on the scaling form of Hamiltonian walks on fractal lattices {\it J. Phys.: Conf. Ser.} {\bf 1814} 012005




    \bibitem{Owc} Owczarek~A~L  and  Prellberg~T 2001 Scaling of self-avoiding walks in high dimensions {\it J. Phys. A: Math. Gen.} {\bf 34}  5773\textendash80


\bibitem{Dus3} Lekić~D and  Elezović-Hadžić~S 2011  Semi-flexible compact polymers on fractal lattices {\it Physica A } {\bf 390}  1941\textendash52

\bibitem{Dus4}  Marčetić~D,  Elezović-Hadžić~S,  Ad\v{z}i\'{c}~N  and \v{Z}ivi\'{c}~I 2019 Semi-flexible compact polymers in two dimensional nonhomogeneous confinement {\it J. Phys. A: Math. Theor. 52} 125001













\end{thebibliography}
\end{document}